\documentclass{llncs}
\usepackage{llncsdoc}
\usepackage{epsf}
\usepackage{amsmath}

\newcommand\nc{\newcommand}

\def\qed{\rule{1.5mm}{3mm}}

\nc{\crl}[2]{\begin{corollary}\label{crl:#1} #2 \end{corollary}}
\nc{\lem}[2]{\begin{lemma}\label{lem:#1} #2 \end{lemma}}
\nc{\prp}[2]{\begin{proposition}\label{prp:#1} #2 \end{proposition}}
\nc{\thm}[2]{\begin{theorem}\label{thm:#1} #2 \end{theorem}}
\nc{\cnj}[2]{\begin{conjecture}\label{cnj:#1} #2 \end{conjecture}}
\nc{\que}[2]{\begin{question}\label{que:#1} #2 \end{question}}
\nc{\pro}[2]{\begin{problem}\label{pro:#1} #2 \end{problem}}
\nc{\dfn}[2]{\begin{definition}\label{def:#1} #2 \end{definition}}
\nc{\rul}[2]{\begin{rulee}\label{rul:#1} #2 \end{rulee}}

\nc{\fig}[4]{\begin{figure}[hbt]
\centerline{
\epsfysize=#2in
\epsffile{#4}
}
\caption{#3}
\label{fig:#1}
\end{figure}}

\nc{\tbl}[3]{\begin{table}[hbt]
#3
\caption{#2}
\label{tab:#1}
\end{table}}

\nc{\refr}[1]{Rule~\ref{rul:#1}}
\nc{\refc}[1]{Corollary~\ref{crl:#1}}
\nc{\reff}[1]{Figure~\ref{fig:#1}}
\nc{\refj}[1]{Conjecture~\ref{cnj:#1}}
\nc{\refl}[1]{Lemma~\ref{lem:#1}}
\nc{\refp}[1]{Proposition~\ref{prp:#1}}
\nc{\reft}[1]{Theorem~\ref{thm:#1}}
\nc{\refq}[1]{Question~\ref{que:#1}}
\nc{\refb}[1]{Problem~\ref{pro:#1}}
\nc{\reffct}[1]{Fact~\ref{fct:#1}}
\nc{\reftb}[1]{Table~\ref{tab:#1}}

\begin{document}

\title{
\Large \bf A Note on Finding Dual Feedback Vertex Set
}
\author{Junjie Ye}
\institute{Department of Computer Science and Engineering, The Chinese University of Hong Kong, Shatin, Hong Kong SAR, China \\
\email{jjye@cse.cuhk.edu.hk}
}
\maketitle

\begin{abstract}
For an edge-bicolored graph $G$ where each edge is colored either red or blue, 
a vertex set $S$ is a dual feedback vertex set 
if $S$ hits all blue cycles and red cycles of $G$. 
In this paper, we show that a dual feedback vertex set of size at most $k$ can be found in time $O^*(c_1^k)$ 
and all minimal dual feedback vertex set of size at most $k$ can be enumerated 
in time $O^*(c_2^{k^2 + k})$ by compact representations for constants $c_1$ and $c_2$. \\

\end{abstract}

\section{Introduction}

Given an undirected graph $G$, the {\sc Feedback Vertex Set} problem (FVS) asks 
if there is a vertex set of size at most $k$ that hits all cycles of $G$. 
FVS is one of the most fundamental graph problems 
whose NP-completeness is proved by Karp~\cite{karp72}.
FVS has been extensively studied under the framework of parameterized complexity~\cite{downey1999parameterized}. 
Since Downey and Fellows~\cite{downey1992fixed}, and Bodlaender~\cite{bodlaender1994disjoint} gave the first FPT algorithms of FVS, there is a long race for the fastest FPT algorithm~\cite{downey1999parameterized,raman2002faster,kanj2004parameterized,raman2006faster,guoa2006compression,dehne20072o,chen2008improved,cao2010feedback,kociumaka2014faster}. 
Besides the classical feedback vertex set problem, many interesting generalizations of FVS have been studied. 
For example, {\sc Subset Feedback Vertex Set}, which requires to hit all cycles going through some specified vertices, 
is shown to be FPT~\cite{wahlstrom2014half,lokshtanov2015linear}.
{\sc Group Feedback Vertex Set}, a more generalized problem, 
is studied and proved to be FPT by Cygan et al.~\cite{cygan2012group}, 
where each edge is labeled with some group element, 
and it requires to hit all cycles that evaluate to a non-null element. 

In this paper, we focus on edge-bicolored graphs 
that are simple undirected graphs with each edge colored either blue or red, 
and consider the parameterized complexity of the following generalization of FVS:  
\begin{quote}
	{\sc Dual Feedback Vertex Set} (DFVS): 
	
	{\sc Input}: Edge-bicolored graph $G$, parameter $k$.
	
	{\sc Question}: Find a vertex set $S$ of size at most $k$ that hits all blue and red cycles. 
\end{quote}

Recently, Cai and Ye~\cite{cai2014dual} have systematically considered such kind of vertex deletion problems on edge-bicolored graphs, 
i.e., problems of deleting vertices on an edge-bicolored graph to make the blue and red graphs satisfy some properties respectively. 
They give a complete characterization of the {\sc Induced $(\Pi_b, \Pi_r)$-Subgraphs} problems for hereditary properties $\Pi_b$ and $\Pi_r$, 
which finds, on an edge-bicolored graph, an induced subgraph of $k$ vertices 
whose blue and red graphs satisfy the properties $\Pi_b$ and $\Pi_r$ respectively. 
Their results include the W[1]-hardness of the parametric dual of DFVS, 
i.e., {\sc Induced $(\Pi_b, \Pi_r)$-Subgraphs} for $\Pi_b = \Pi_r$ being acyclic. 

In this paper, we give an FPT algorithm for DFVS, 
and show that all minimal solutions of size at most $k$ can be enumerated in FPT time by compact representations. 
In connection with these results, we generalize the FPT algorithm of DFVS to show that, given an edge-colored graph $G$ with $h$ colors, 
it takes time $O^*(c^{hk\log k})$ \footnote{ The $O^*$-notation suppresses factors polynomial in the input size. } 
to find a vertex set of size at most $k$ that hits all monochromatic cycles of $G$. 

For a graph $G$, $V(G)$ and $E(G)$ denote its vertex set and edge set respectively,
and $n$ and $m$, respectively, are numbers of vertices and edges of $G$.
For a subset $V' \subseteq V(G)$, $N_G(V')$ denotes the neighbors of $V'$ in $V(G) - V'$
and $G[V']$ the subgraph of $G$ induced by $V'$.

For an edge-bicolored graph $G = (V, E_b \cup E_r)$,
$G_b = (V, E_b)$ and $G_r = (V, E_r)$, respectively, denote the blue graph and red graph of $G$. 
For a vertex $v \in V$, $d_b(v)$ and $d_r(v)$ denote the degree of $v$ in $G_b$ and $G_r$ respectively. 
We use $d(v) = (d_1, d_2)$ to denote the degree of $v$ in $G$, where $d_1 = d_b(v)$ and $d_2 = d_r(v)$.
A subset $V' \subseteq V(G)$ is a {\em dual feedback vertex set} of $G$
if both $G_b - V'$ and $G_r - V'$ are acyclic.

An instance $(I, k)$ of a parameterized problem $\Pi$ consists of two parts: an input $I$ and a parameter $k$. 
We say that a parameterized problem $\Pi$ is fixed parameter tractable (FPT) if there is an algorithm solving every instance $(I, k)$
in time $f(k)|I|^{O(1)}$ for some computable function $f$.

In the rest of the paper, we present FPT algorithms for DFVS in Section 2, 
and show how to enumerate minimal solutions of DFVS in Section 3. 
We conclude with discussions in Section 4.

\section{Dual Feedback Vertex Set}
In this section, we show that DFVS can be solved in time $O^*(c^k)$ 
following the result of Guo et al.'s~\cite{guoa2006compression}
which enumerates all minimal feedback vertex sets in FPT time by compact representations. 
A compact representation $C$ is a collection of pairwise disjoint vertex sets. 
A vertex set $S$ that contains exactly one element of every set
of a compact representation $C$ is a minimal dual feedback set of size at most $k$, 
and we say that $S$ is represented by $C$. 

\thm{compact}{
DFVS can be solved in time $O^*(c^k)$ for a constant $c$.
}
\proof{
We first enumerate all compact representations 
of minimal feedback vertex set of size at most $k$ for both $G_b$ and $G_r$ 
in time $O^*(c_1^k)$ for a constant $c_1$ by Guo et al.'s algorithm~\cite{guoa2006compression}. 
Since a minimal dual feedback vertex set is always a union of 
a minimal feedback vertex set of $G_b$ and a minimal feedback vertex set of $G_b$, 
a minimal solution of DFVS is a minimal vertex set that 
covers every vertex set in $C_b \cup C_r$ for two compact representations $C_b$ of $G_b$ and $C_r$ of $G_r$. 
Given two compact representations $C_b$ and $C_r$ of $G_b$ and $G_r$ respectively, 
we show that a minimum vertex set $S$ that intersects every vertex set in $C_b \cup C_r$ can be found in polynomial time. 

To this end, we build an auxiliary graph $H$ as following: 
\begin{itemize}
\item For a vertex set $S_i \in C_b \cup C_r $, create a vertex $s_i$. 

\item If a vertex $v$ is in both $S_i \in C_b$ and $S_j \in C_r$, add an edge $e_v = s_is_j$. 

\item If there are $t$ parallel edges between a pair of vertices, 
select one edge to represent the $t$ edges and delete other $t - 1$ edges. 
\end{itemize}
Note that each edge represents the intersection of two sets, 
thus if there are $t$ parallel edges we can arbitrary delete $t - 1$ edges. 
Since $C_b$, also $C_r$, contains at most $k$ vertex disjoint sets, 
$H$ has at most $2k$ vertices and $k^2$ edges. 
For each isolated vertex $s_i$ of $H$, we arbitrary put a vertex $v \in S_i$ into $S$. 
Then we find a minimum edge set $F$ that covers all non-isolated vertices, 
which can be done in time $O(k^3)$ following the Theorem 3.7 in Cai's paper~\cite{cai2008parameterized}. 
For an edge $e_v \in F$, put the vertex $v$ into $S$. 
Then $S$ is a minimum vertex set that covers all sets of $C_b \cup C_r$. 

As there are $c_1^k$ ways to select a compact representation $C_b$, also $C_r$, 
the running time for this algorithm is $O^*(c_1^{2k}) = O^*(c^k)$ for a constant $c$. 
\qed} \\

One crucial step of the above algorithm is to find a minimal set that covers two compact representations. 
We observe that the analogous step can be done in FPT time for {\sc Multi-Feedback Vertex Set} (MFVS)
which is the problem of finding a vertex set of size at most $k$ 
that hits all monochromatic cycles on an edge-colored graph. 
Based on the observation, we can establish the fixed parameter tractability of MFVS. 

\thm{}{
	MFVS can be solved in time $O^*(c^{hk\log k})$ where $h$ is the number of colors of the input edge-colored graph. 
}
\proof{
	Let $G$ be an edge-colored graph with $h$ colors, and $G_1, \dots, G_h$ be the monochromatic subgraphs of $G$. 
	We first obtain compact representations of minimal feedback vertex set of size at most $k$ for $G_i$ 
	by the algorithm of Guo et al.'s~\cite{guoa2006compression}, which takes time $O^*(hc_1^k)$ for a constant $c_1$. 
	
	Let $C_i$ be a compact representation of $G_i$. 
	Since a minimal solution of {\sc Multi-Feedback Vertex Set} is always a union of minimal feedback sets of $G_i$, 
	a vertex set of size at most $k$ that intersects all sets of $C_i$ for $1 \le i \le h$ will be a solution. 
	To find such a vertex set, we build an auxiliary graph $H$ as following:
	\begin{itemize}
		\item For each vertex set $S_i \in C_1 \cup \dots \cup C_h$, 
		create a {\em set-vertex} $s_i$. 
		
		\item For each vertex $v$ of $G$, create a {\em copy-vertex} $v'$ and 
		add edge $v's_i$ if $v \in S_i$. 
		
		\item Create an edge $v^*u^*$, and add all edges between $v^*$ and copy-vertices.
	\end{itemize}
	It is easy to check that $H$ has a dominating set of size at most $k + 1$ 
	iff there is a set of at most $k$ copy-vertices that dominates all set-vertices, 
	and thus corresponds to a solution of MFVS. 
	Furthermore, given a dominating set $S$ of size at most $k + 1$ of $H$, 
	we can obtain a set of copy-vertices which corresponds to a solution of MFVS 
	by replacing each set-vertex $s_i \in S - \{v^*, u^*\}$ by a copy-vertex in $N(s_i)$.  
	
	Since the vertex sets are pairwise disjoint in a compact representation, 
	each copy-vertex is adjacent to at most one set-vertex of a compact representation. 
	Each copy-vertex has degree at most $h + 1$, and thus $H$ is a $(h+1)$-degenerated bipartite graph. 
	Then we can apply the algorithm of Alon and Gutner~\cite{alon2009linear}, 
	which finds a dominating set of size at most $l$ on a $d$-degenerated graph in time $l^{O(dl)}n$, 
	to obtain a dominating set of size at most $k + 1$ in time 
	$(k+1)^{O((h + 1)(k + 1))}n = O(c_2^{hk\log k}n)$ for a constant $c_2$. 
	Therefore, given one compact representation of each $G_i$, 
	we can find a solution among these compact representations in time $O(c_2^{hk\log k}n)$. 
	
	Since there are $c_1^k$ ways to choose a compact representation $C_i$ of $G_i$, 
	the total running time of the algorithm is 
	$O^*(c_1^{hk} c_2^{hk\log k}) = O^*(c^{hk\log k})$ for a constant $c$.
\qed}

\section{Dual Feedback Vertex Set Enumeration}
In this section we give two FPT algorithms to enumerate all minimal dual feedback vertex set 
of size at most $k$ by compact representations. 

We first introduce the algorithm based on the proof of \reft{compact}. 
In the auxiliary graph $H$ of \reft{compact}, a vertex $s_i$ represents a set $S_i$ 
and an edge $s_is_j$ represents the vertex set $S_i \cap S_j$. 
We need to cover all vertices of $H$ by selecting at most $k$ edges or vertices. 
Since $H$ contains at most $2k$ vertices and $k^2$ edges, 
we can enumerate all minimal sets of at most $k$ elements 
that covers all vertices of $H$ in time $O^*(2^{k^2 + 2k})$ by exhausted search. 
Each minimal set is a compact representation of minimal dual feedback vertex sets 
of the original edge-bicolored graph $G$. 
And the running time is $O^*(2^{k^2 + 2k} c_1^{2k}) = O^*(c^{k^2 + k})$ for a constant $c$. 

Next we present another FPT algorithm which follows the idea of iterative compression 
introduced by Reed, Smith and Vetta~\cite{reed2004finding}. 
The main portion is to solve the following problem: 
\pro{problem}{
Given an edge-bicolored graph $G = (V, E_b \cup E_r)$ and a dual feedback vertex set $S$, 
find a dual feedback vertex set $S'$ such that $S' \cap S = \emptyset$ and $|S'| < |S|$.
}

We show that all minimal solutions of Problem 1 can be enumerated by compact representations in FPT time. 
To this end, we introduce some simple reduction rules.

\rul{degree-0}{
If there is a vertex $v$ with $d(v) = (0, 0)$, reduce to $G = G - v$.
}

\rul{degree-1}{
If there is a vertex $v$ with $d_b(v) = 1$ (resp., $d_r(v) = 1$), 
remove the blue (resp., red) edge incident to $v$.
}

\rul{degree-2}{
For vertices $v \notin S$ and $x, y \in S$ 
such that $x$ and $y$ belong to the same blue (resp., red) component of $G[S]$, 
and there are two blue (resp., red) edges $vx$ and $vy$, 
reduce to $G = G - v$ and add $v$ to $S'$.
}

Above three rules are obviously safe, and can be executed in polynomial time. 
Call an edge-bicolored graph {\em reduced graph} if none of above rules is applicable. 
Then in a reduced graph, there are no vertex $v \notin S$ such that $d(v) = (0, 0)$, $(1, *)$ or $(*, 1)$. 
In a reduced graph $G$, let $V^b_{> 2}$ (resp., $V^r_{>2}$) be the set of vertices in $V(G) - S$ whose blue (resp., red) degree is larger than 2, 
$V_{\le 2}$ be the set of vertices whose both blue degree and red degree are at most 2, i.e., $V_{\le 2} = V(G) - S - V^b_{> 2} - V^r_{> 2}$. 
A blue path $P = \{v_1, v_2, \dots, v_p\}$ is a {\em maximal blue path} 
if $v_i \notin S$, $d_b(v_i) = 2$ for $1 \le i \le p$ and 
both $v_1$ and $v_p$ are adjacent to vertices in $S \cup V^b_{>2}$. 
A {\em maximal red path} is defined similarly. 

Now we introduce two rules that reduce the size of a maximal blue (red) path in a reduced graph. 

\rul{path-1}{
Let $P_b$ (resp., $P_r$) be a maximal blue (resp., red) path, 
and $V^*$ be the vertex set containing all vertices of $P_b$ (resp., $P_r$) with degree $(2, 0)$ (resp., $(0, 2)$). 
If $|V^*| \ge 2$, select one vertex $v \in V^*$ to represent $V^*$. 
For other vertices $u \in V^*$, remove $u$ and 
add a blue (resp., red) edge between $u$'s two blue (resp., red) neighbors. 
}

\rul{path-2}{
Let $P_b$ and $P_r$ be a maximal blue path and red path respectively. 
If $P_b$ and $P_r$ share common vertices, 
select one common vertex $v$ to represent all common vertices, i.e., $V(P_b) \cap V(P_r)$. 
For other vertices $u \in V(P_b) \cap V(P_r)$, remove $u$ and 
add a blue (resp., red) edge between $u$'s two blue (resp., red) neighbors. 
}

\lem{path}{
\refr{path-1} and \refr{path-2} are safe. 
Let $m_b$ and $m_r$, respectively, be the numbers of maximal blue paths and maximal red paths. 
After applying \refr{path-1} and \refr{path-2}, $|V_{\le 2}| \le (m_b + 1)*(m_r + 1)$. 
}
\proof{
Suppose that $v$ and $u$ are both on a maximal blue path $P_b$ and a maximal red path $P_r$. 
Since all monochromatic cycles going through vertex $u$ also go through $v$, 
if a solution contains $u$, we can replace $u$ with $v$. 
A minimal solution may contain at most one vertex from $V(P_b) \cap V(P_r)$, 
and if it does contain one such vertex, it doesn't matter which vertex it contains. 
Thus we can use one arbitrary vertex to represent $V(P_b) \cap V(P_r)$ and dissolve other vertices. 
Therefore, \refr{path-2} is safe. Similarly, \refr{path-1} is safe. 
Note that \refr{path-1} and \refr{path-2} can be executed in polynomial time. 

Now consider a maximal blue path $P_b$. 
We bound the number of vertices in $V(P_b) \cap V_{\le 2}$. 
By \refr{path-1}, $P_b$ contains at most one vertex with degree $(2, 0)$. 
By \refr{degree-1}, $P_b$ contains no vertices with degree $(2, 1)$. 
Then we consider vertices of $P_b$ with degree $(2, 2)$.
Note that each such vertex is an intersection of $P_b$ and a maximal red path. 
By \refr{path-2}, the number of such vertices is at most $m_r$. 
Therefore $|V(P_b) \cap V_{\le 2}| \le 1 + m_r$. 
Similarly, for a maximal red path $P_r$, $|V(P_r) \cap V_{\le 2}| \le 1 + m_b$. 
Since each vertex in $V_{\le 2}$ is on a maximal blue (red) path, 
we have $|V_{\le 2}| \le (m_b + 1)*(m_r + 1)$. 
\qed} \\ 

Let us focus on the blue (red) subgraph of an reduced graph. 
Guo et al. show that there are at most $14|S|$ vertices with degree larger than 2 
and at most $16|S|$ maximal blue (red) paths in the proof of Lemma 2 and Lemma 6 in~\cite{guoa2006compression}.
Relying on these results, we enumerate all minimal solutions of \refb{problem} in FPT time. 

\lem{problem}{
All minimal solutions of \refb{problem} can be enumerated in $O^*( c^{|S|^2 + |S|} )$ time for a constant $c$ 
by compact representations. 
}
\proof{
Reduce the input edge-bicolored graph $G$ by \refr{degree-0} to \refr{path-2}. 

We first determine which vertices of $V_{>2} = V^b_{> 2} \cup V^r_{> 2}$ of $G$ are in a minimal solution.
By the proof of Lemma 2 in~\cite{guoa2006compression}, we have $|V_{>2}| \le 28|S|$. 
We guess which vertices of $V_{>2}$ are in a solution $S'$ and branch into $\sum\limits_{t = 0}^{|S| - 1} \binom{28|S|}{t}$ cases. 
In each case, we put $t$ one-element sets into $C$, 
delete these vertices in $G$ and apply \refr{degree-0} to \refr{path-2} to further reduce the graph.
Denote by $G'$ the resulting new edge-bicolored graph . 

Then we determine, for each case, which vertices of $V_{\le 2}$ of $G'$ are in a minimal solution.
Note that every vertex, that is in $V_{\le 2}$ and also remains in $G'$, 
is on a maximal blue (red) path of $G'$. 
By the proof of Lemma 6 in~\cite{guoa2006compression}, 
there are at most $16|S|$ maximal blue paths, also maximal red paths in $G'$. 
According to \refl{path}, $|V_{\le 2}| \le (16|S| + 1)^2$,
and we can use exhausted search to find vertices in $V_{\le 2} \cap S'$. 

The total running time is $O^*(\sum\limits_{t = 0}^{|S| - 1} \binom{28|S| + (16|S|)^2}{t}) = O^*(c^{|S|^2 + |S|})$ for a constant $c$. 
\qed} \\

Obtaining \refl{problem}, we are capable to enumerate all minimal solutions of DFVS by compact representations. 

\thm{dual-FVS}{
All minimal solutions of DFVS can be enumerated in time $O^*(c^{k^2 + k})$  
by compact representations for a constant $c$. 
}
\proof{
We first find feedback vertex sets $S_b$ and $S_r$ of size at most $2k$ for $G_b$ and $G_r$ respectively, 
by the approximation algorithm of Bafna et al.~\cite{bafna19992}. 
Then $X = S_b \cup S_r$ is a dual feedback vertex set of size at most $4k$. 
Let $X'$ be a minimal solution of size at most $k$. 
We guess which vertices of $X$ are in $X'$, i.e., the vertex set $Y = X \cap X'$, 
and branch into $\sum\limits_{|Y| = 0}^{k} \binom{|X|}{|Y|}$ cases.
For each case, $X - Y$ is a dual feedback vertex set of $G - Y$ 
and $X' - Y$ is a smaller dual feedback vertex set that is disjoint from $X - Y$. 
By \refl{problem}, all possible sets $X' - Y$ can be enumerated in time $O^*(c_1^{(|X|-|Y|)^2 + |X|-|Y|})$. 
Then combining $Y$ and $X' - Y$, we get all minimal solutions of size at most $k$. 
The total running time for the algorithm is $O^*(\sum\limits_{|Y| = 0}^{k} \binom{|X|}{|Y|} c_1^{(|X|-|Y|)^2 + |X|-|Y|})$ 
which is $O^*(c^{k^2 + k})$ for a constant $c$ as $|X| \le 4k$ and $|Y| \le k$. 
\qed} 

\section{Concluding remarks}
In this paper, we established the fixed parameter tractability of DFVS, 
and provided two algorithms to enumerate all minimal solutions of DFVS by compact representations in FPT time. 
Furthermore, we established the fixed parameter tractability of MFVS by generalizing the algorithm of DFVS. 

Note that there are at most $hk$ set-vertices in the auxiliary graph $H$ in the proof of Theorem 2, 
and each copy-vertex dominates at most $h$ set-vertices. 
Thus there are at most $\binom{hk}{h}$ copy-vertices with different neighbors, 
which means that the number of copy-vertices can be up bounded by $\binom{hk}{h}$. 
Then we can use exhausted search of these copy-vertices to enumerate all minimal solutions of MFVS. 
\thm{}{
	All minimal solutions of MFVS can be enumerated in FPT time by compact representations. 
}

For edge-bicolored graphs, instead of considering monochromatic cycles, 
we can study alternating cycles which are cycles alternates between blue and red edges. 
As suggested by Cai Leizhen, it is very interesting to consider the parameterzied complexity of the following problem: 
\begin{quote}
	{\sc Feedback Vertex Set for Alternating Cycles} (AFVS): 
	
	{\sc Input}: Edge-bicolored graph $G$, parameter $k$.
	
	{\sc Question}: Find a vertex set $S$ of size at most $k$ that hits all alternating cycles.
\end{quote}
We observe that AFVS is a generalization of {\sc Directed Feedback Vertex Set}~\cite{chen2008fixed}
which hits all cycles of a digraph. 
Given a digraph $G$, we can obtain an edge-bicolored graph $G'$ 
by replacing each arc $uv$ by a blue edge $ux$ and a red edge $xv$. 
It is easy to check that $G'$ has a feedback vertex set of size at most $k$ for alternating cycles 
iff $G$ has a directed feedback vertex set of size at most $k$. \\

{\noindent \bf Acknowledgments.} The author is grateful to Cai Leizhen for helpful discussions and suggestions. 

\bibliographystyle{splncs03}
\bibliography{DFVS}

\end{document}